\documentclass[12pt]{article}
\usepackage{amsmath, amsfonts, bm, mathtools, algpseudocode, amssymb}
\usepackage[dvips]{graphicx}
\usepackage{psfrag,epsf,algorithm}
\usepackage{enumerate}
\usepackage{natbib}
\usepackage{url} 

\newcommand{\blind}{0}

\usepackage[utf8]{inputenc}
\usepackage[english]{babel} 
\newtheorem{theorem}{Theorem}[section]

\newtheorem{lemma}[theorem]{Lemma}

\begin{document}

\def\spacingset#1{\renewcommand{\baselinestretch}%
{#1}\small\normalsize} \spacingset{1}


\if0\blind
{
  \title{\bf A Factor Stochastic Volatility Model with Markov-Switching Panic Regimes}
  \author{Taylor R. Brown \hspace{.2cm}\\
    Department of Statistics, University of Virginia}
  \maketitle
} \fi

\if1\blind
{
  \bigskip
  \bigskip
  \bigskip
  \begin{center}
    {\LARGE\bf Factor Stochastic Volatility Models with Markov-Switching Panic Regimes}
\end{center}
  \medskip
} \fi

\bigskip
\begin{abstract}
The use of factor stochastic volatility models requires choosing the number of latent factors used to describe the dynamics of the financial returns process; however, empirical evidence suggests that the number and makeup of pertinent factors is time-varying and economically situational. We present a novel factor stochastic volatility model that allows for random subsets of assets to have their members experience non-market-wide panics. These participating assets will experience an increase in their variances and within-group covariances. We also give an estimation algorithm for this model that takes advantage of recent results on Particle Markov chain Monte Carlo techniques.
\end{abstract}

\noindent%
\vfill

\newpage
\spacingset{1.45} 
\section{Introduction}
\label{sec:intro}

Models for multivariate financial time series are used to help manage investment risk and choose portfolios in the presence of time-varying means and covariances. Among the models suitable for this task, \emph{factor stochastic volatility} (FSV) models possess a low-dimensional latent process upon which the distribution of the returns depends \cite{Chib2009}. Typically, an element of this latent process will represent volatility (or log volatility). As this process rises (or falls), the scale of the return distribution will rise (or fall) \cite{aguilarwest}, \citep{jacquier1999}. These processes may also represent ``regimes" of returns; if this is the case, the process will have a finite state space, and conditional on its value at time $t$, one or more aspects of the return distribution will change \cite{hamilton}.

FSV models are able to forecast effectively because they can be quite expressive; however, there are two main difficulties associated with their use. For one, model specification will always be up for debate. Pertinent unobserved quantities that vary over time must be included in the state process, and ones that do not are to be included as parameters. Choosing relevant quantities and separating them into these two groups is just the beginning of the model specification task, and even if the correct choice is made, quantities that are assumed constant can eventually change abruptly in the future data generating process. This may explain why, in practice, a model that forecasts well for a time might lose its forecasting ability. 

The second difficulty is computation, particularly in the task of model estimation. FSV models are part of a much broader category of time series models called state-space models. In this category, it is often the case that the marginal likelihood cannot be evaluated, owing to the fact that it is a high dimensional integral. If one adopts a Bayesian approach, Gibbs sampling or Metropolis-Hastings algorithms are available that sidestep this difficulty; however, these algorithms are difficult to tune, can be model-dependent, and might restrict the selection of available prior distributions \cite{asis}, \cite{pitt1999time}. These difficulties render stochastic volatility models unfeasible for many practitioners.

This paper addresses these two difficulties by proposing a new Markov-switching factor stochastic volatility model and by making use of recent advances in particle Markov chain Monte Carlo methods to develop a highly general and extensible estimation procedure. The model attempts to forecast volatility more scrupulously, instead of just adapting to it. In attempting this, the model is able to signal which of several assets are experiencing a non-market-wide, ``contained" panic. The estimation procedure is a variant of the Metropolis-Hastings algorithm \cite{mh}, which is interesting in its own right, as these types of algorithms are rarely applied to FSV models. 

The structure of this paper is as follows. First, the model is described in detail in section 2. Then, in section 3, an overview of particle filtering is given, along with the details of our version of a particle marginal Metropolis-Hastings algorithm. Section 4 estimates a versions of the model, and uses the estimated parameters to construct portfolios and estimate risk measures on out-of-sample data. Finally, a short conclusion is given.



\section{Methods}

\subsection{Model Description} \label{sec:modeldescription}

Let $\mathbf{y}_t \in \mathbb{R}^{d_y}$ denote the returns at time $t$ for $d_y$ assets. The observation equation describes the distribution of the returns conditional on contemporaneous factors. This equation can be written as:
\begin{eqnarray} \label{obseqn}
\mathbf{y}_t = \mathbf{B}\mathbf{f}_{1,t} + \mathbf{D}({x_{1,t}})\mathbf{B}\mathbf{f}_{2,t} + \mathbf{v}_t
\end{eqnarray}
where $\mathbf{B} \in \mathbb{R}^{d_y \times d_f}$ is a loadings matrix whose elements are analogous to the coefficients of a standard linear regression model, $\{\mathbf{v}_t\} \overset{\text{iid}}{\sim} \text{Normal}(\mathbf{0}, \mathbf{R})$ are error disturbances, $\{x_{1,t}\} \in \mathbb{N}^+$ is a finite state space Markov chain, $\mathbf{D} : \mathbb{N}^+ \to \mathbb{R}^{d_y \times d_y}$ is a function that yields a ``selector" matrix (whose explanation is forthcoming), and $\mathbf{f}_t= (\mathbf{f}_{1,t}^\intercal, \mathbf{f}_{2,t}^\intercal)^\intercal $ is the latent factor vector defined as:
\begin{eqnarray} \label{new_factor}
\mathbf{f}_t =
\left[\begin{array}{c}
\bm{\lambda_1} \\
\bm{0}
\end{array}\right] +
\left[\begin{array}{cc}
\exp(\text{diag}[\mathbf{x}_{2t}]/2) & 0 \\
0 & \exp(\text{diag}[\mathbf{x}_{3t}]/2)
\end{array}\right]
\left[\begin{array}{c}
\mathbf{z}_{1,t}\\
\mathbf{z}_{2,t}
\end{array}\right].
\end{eqnarray}
Here $\bm{\lambda}_1$ is the factor ``risk premia" vector, $\exp(\cdot)$ is the matrix exponential, and $\{\mathbf{z}_{2,t}\}$ and $\{\mathbf{z}_{2,t}\}$ are iid standard Gaussian noise vectors.

Two subvectors of the state, $\mathbf{x}_{2,t},\mathbf{x}_{3,t} \in \mathbb{R}^{d_f}$, are vectors of latent log volatilities. They control the scale of both factor vectors, and we assume all of their elements evolve as independent and stationary AR(1) processes:

\begin{align*} \label{fsvstate}
\left[\begin{array}{c}
\mathbf{x}_{2,t} \\
\mathbf{x}_{3,t}
\end{array}\right]  &= 
\left[\begin{array}{c}
\bm{\mu}_1 \\
\bm{\mu}_2
\end{array}\right]
+ \left[\begin{array}{cc}
\bm{\Phi}_1 & 0 \\
0 & \bm{\Phi}_2
\end{array}\right]
\left( 
\left[\begin{array}{c}
\mathbf{x}_{2,t-1} \\
\mathbf{x}_{3,t-1}
\end{array}\right]
- 
\left[\begin{array}{c}
\bm{\mu}_1 \\
\bm{\mu}_2
\end{array}\right]
\right) + 
\left[\begin{array}{c}
\mathbf{w}_{1,t} \\
\mathbf{w}_{2,t}
\end{array}\right]
, \\
\left[\begin{array}{c}
\mathbf{x}_{2,1} \\
\mathbf{x}_{3,1}
\end{array}\right] 
&\sim 
\text{N}\left(\left[\begin{array}{c}
\bm{\mu}_1 \\
\bm{\mu}_2
\end{array}\right]
, \operatorname{diag}\left[ \frac{\sigma_1^2}{(1-\phi_1^2)}, \ldots, \frac{\sigma_{2d_f}^2}{(1-\phi_{2d_f}^2)} \right] \right),
\end{align*}
where $\{\mathbf{w}_t\} \overset{\text{iid}}{\sim} \text{Normal}(\mathbf{0}, \mathbf{Q})$, $\mathbf{Q} = \text{diag}\left[\begin{array}{cc} \mathbf{Q}_1 & \mathbf{Q}_2 \end{array}\right] = \text{diag}(\sigma_1^2, \ldots, \sigma^2_{2d_f})$, $\bm{\mu}$ is the average log volatility, and $\bm{\Phi}= \text{diag}\left[\begin{array}{cc} \bm{\Phi}_1 & \bm{\Phi}_2 \end{array}\right] =  \text{diag}(\phi_1, \ldots, \phi_{2d_f})$ is the state transition matrix. 

The random selector matrix $\mathbf{D}(x_{1,t})$ is diagonal with entries that are either $0$ or $1$. If a $1$ is placed in the $(i,i)$th position, this means that the $i$th return is participating in a contained panic.  The placement of these $1$s is random, and controlled by the process $x_{1,t}$. Notice that, if all the elements of $\mathbf{D}(x_{1,t})$ are zero, then the observation equation resembles that of a more simplistic FSV model. 

For simplicity, it is assumed that $x_{1,t}$ evolves independently of the other state processes. Furthermore, the size of the state space of $x_{1,t}$ is dependent on a user-chosen number $K < d_y$, which denotes the maximum number of assets that can be affected by a contained panic. $K$ will always be chosen to be less than $d_y$ so that $\mathbf{f}_{2,t}$ will affect at most $K$ elements of the return vector, and retain its interpretation as a contained panic factor. Define $S_K = \sum_{k=0}^K {d_y \choose k}$, which is the number of possible subsets of financial assets that are at most size $K$. 

The state space for $x_{1,t}$ is set to $\{1, 2, \ldots, S_K\}$. These numbers enumerate the lexicographically ordered possible diagonals of $\mathbf{D}(x_{1,t})$. For example, if $d_y = 3$, and $K=2$, then the state space of $x_{1,t}$ is $\{1, 2, \ldots, 7\}$, and these numbers label the possible values of $\mathbf{D}(x_{1,t})$ which are:
\[\left\{\text{diag}[0,0,0], \text{diag}[0,0,1], \text{diag}[0,1,0], \text{diag}[0,1,1], \text{diag}[1,0,0], \text{diag}[1,0,1], \text{diag}[1,1,0] \right\}.
\]

The most common regime will likely be the first, which zeros out the second term of equation (\ref{obseqn}). However, when the $(i,i)$th element of $\mathbf{D}(x_{1,t})$ is $1$, then the $i$th row of $\mathbf{D}(x_{1,t}) \mathbf{B}$ will be nonzero, and there will be an effect on stock $i$'s variance. There will also be an effect on the covariance between stock $i$ and any other stock corresponding to a nonzero row of $\mathbf{D}(x_{1,t}) \mathbf{B}$.

The transition matrix is parameterized by a single parameter: $0 < p < 1$. It is written as:
\begin{eqnarray}
p \mathbf{I}_{S_K} + \frac{1-p}{S_K - 1}(\mathbf{1}_{S_K}\mathbf{1}_{S_K}^T - \mathbf{I}_{S_K}).
\end{eqnarray}
The associated chain is assumed to start in its stationary distribution, which is the discrete uniform distribution.

This model is not identifiable without restrictions on the parameter space. There are nonidentifiabilities linked to scaling, permutation, and sign switching in the columns of the loadings matrix. Proofs for the following lemmata can be found in the appendix.

\begin{lemma}
The MSL model defined by the collection of parameters $\{\mathbf{B}, \mathbf{R}, \bm{\mu}, \bm{\Phi}, \mathbf{Q}, \bm{\lambda}, p \}$ yields the same likelihood as the MSL model with $\{\mathbf{BP}, \mathbf{R}, \bm{\mu} +\bm{u}, \bm{\Phi}, \mathbf{Q}, \bm{P}^{-1}\bm{\lambda}, p\}$, where $\mathbf{P}$ is a diagonal matrix with all its entries positive, $\log(\cdot)$ denotes the matrix logarithm obtained by taking the log of all the diagonal elements, and 
\[
\bm{u}^{\intercal} = (2\log(\bm{P}^{-1})_{11}, \ldots, 2\log(\bm{P}^{-1})_{nn}).
\]
\end{lemma}

\begin{lemma}
The MSL model defined by the collection of parameters $\{\mathbf{B}, \mathbf{R}, \bm{\mu}, \bm{\Phi}, \mathbf{Q}, \bm{\lambda}, p \}$ yields the same likelihood as the MSL model with $\{\mathbf{BP}^{\intercal}, \mathbf{R}, \mathbf{P}\bm{\mu}, \mathbf{P}\bm{\Phi}\mathbf{P}^{\intercal}, \mathbf{P}\mathbf{Q}\mathbf{P}^{\intercal} , \mathbf{P}\bm{\lambda}, \bm{p}\}$, where $\mathbf{P}$ is a permutation matrix.
\end{lemma}

\begin{lemma}
The MSL model defined by the collection of parameters $\{\mathbf{B}, \mathbf{R}, \bm{\mu}, \bm{\Phi}, \mathbf{Q}, \bm{\lambda}, p \}$ yields the same likelihood as the MSL model with $\{\mathbf{BS}, \mathbf{R}, \bm{\mu}, \bm{\Phi}, \mathbf{Q}, \bm{\lambda}, \bm{p} \}$, where $\mathbf{S}$ is a diagonal matrix with all nonzero elements equal to $\pm 1$.
\end{lemma}

To prevent these nonidentifiabilities, we adopt the measures taken in \cite{apt} and assume the loadings matrix is of the form where the $(i,j)$th entry is set to $1$ whenever $i=j$, and $0$ when $i < j$. Whenever $i > j$, this element is a free parameter that needs to be estimated. Also, we restrict $d_f$ so that the number of free parameters in the observation covariance matrix is fewer than $d_y + { d_y \choose 2}$. 

As an example, consider the case where $d_f=1$. From \ref{obseqn}, the formulas for the conditional mean vector and covariance matrix are 
\begin{gather*} \label{meanvar1}
\mathbb{E}[\mathbf{y}_t \mid \mathbf{x}_t, \theta] =  \lambda \mathbf{B} \\
\mathbb{V}[\mathbf{y}_t \mid  \mathbf{x}_t, \theta] = \exp(x_{2,t})\mathbf{B}\mathbf{B}^{\intercal} + \exp(x_{3,t})\mathbf{D}(x_{1,t})\mathbf{B}\mathbf{B}^{\intercal}\mathbf{D}(x_{1,t}) + \mathbf{R}
\end{gather*}
where $\theta = (\mathbf{B}, \mathbf{R}, \bm{\mu}, \bm{\phi}, \mathbf{Q}, \lambda, p)$ is the collection of all model parameters. If the $i$th diagonal of $\mathbf{D}(x_{1,t})$ equals $1$, the variance increases by $\exp(x_{3,t})B_1^2$. If the $(j,j)$th element is $1$ as well, then the covariance between these two elements will change by $\exp(x_{3,t}) B_1B_2$.

\section{Estimation}

Estimation is carried out using the particle marginal Metropolis-Hastings algorithm \cite{pmcmc}. To promote a more thorough understanding of this procedure, particle filtering is described first, using simplified notation. Finally, the specific algorithm used for this model is given, along with a theorem and some guidelines to increase its efficiency.

\subsection{Preliminaries}

Let $(Y, \mathcal{Y})$ and $(X,\mathcal{X})$ be the measurable state spaces of the observed and unobserved time series data, respectively. A state-space model is defined by two things: a $Y$-valued observable time series $\{\mathbf{y}_t\}_{t=1}^T \coloneqq \mathbf{y}_{1:T}$, and an $X$-valued unobservable time series $\{\mathbf{x}_t\}_{t=1}^T \coloneqq \mathbf{x}_{1:T}$. 

Specifying one of these models requires the selection of several probability distributions parameterized by the vector $\theta$. What are chosen are $f(\mathbf{x}_1 \mid \theta)\mu(\text{d}\mathbf{x}_1)$, the first time's state distribution, along with the state transition probabilities, $f(\mathbf{x}_t \mid x_{t-1},\theta) \mu(\text{d}\mathbf{x}_t)$ for $t \ge 2$, and the observation densities, $\{g(\mathbf{y}_{t} \mid \mathbf{x}_{t},\theta)\}_{t=1}^T$. $\mu ( \cdot )$, not to be confused with the average log volatility parameter described in the previous section, is defined as some product of counting and Lebesgue measures. The state space of the model we are interested in will include both continuous and discrete random variables.

A particle filter will yield approximations for both the filtering distributions $p(\mathbf{x}_t \mid \mathbf{y}_{1:t}, \theta), 1 \le t \le T$, and the marginal likelihood $p(\mathbf{y}_{1:T} \mid \theta)$. We use both for out-of-sample forecasting, and the latter to aid in the estimation procedure. At every time point, a particle filter revises its filtering approximation using the following recursions:
\begin{gather*}
p(\mathbf{x}_t \mid \mathbf{y}_{1:t}, \theta) \propto g(\mathbf{y}_t \mid \mathbf{x}_t, \theta) p(\mathbf{x}_t \mid \mathbf{y}_{1:t-1}, \theta) \\
p(\mathbf{x}_t \mid \mathbf{y}_{1:t-1}, \theta) = \int f(\mathbf{x}_t \mid \mathbf{x}_{t-1}, \theta) p(\mathbf{x}_{t-1} \mid \mathbf{y}_{1:t-1},\theta)\mu(\text{d}\mathbf{x}_{t-1}).
\end{gather*}
At time $t-1$, one possesses weighted samples targeting $p(\mathbf{x}_{t-1} \mid \mathbf{y}_{1:t-1}, \theta)$. Then, going into time $t$, he propagates forward his samples and revises each of their weights based on how well they cohere with the new observation $\mathbf{y}_t$. Finally, at the end of every time step, there is the option to resample particles based on these revised weights. 

The most commonly used algorithm, which allows for a wide range of proposal densities used at each time point to propogate particles, is known as sequential importance sampling with resampling (SISR) \cite{tutorial}. To take advantage of our model's strucure, we use a more elaborate ``Rao-Blackwellized" particle filter \cite{chenmixturekalman}. This variance-reduction technique is only possible when the model admits a certain form. 

Assume that there are two levels of the state process, and write $\mathbf{x}_t = (\mathbf{x}_{1,t}^\intercal, \mathbf{x}_{2,t}^\intercal)^\intercal$, where $\mathbf{x}_{1,t}$ and $\mathbf{x}_{2,t}$ are subvectors for time $t$'s state. The necessary form for a model of interest possesses a smoothing distribution that can be written as:
\begin{eqnarray}
p(\mathbf{x}_{1:t} \mid \mathbf{y}_{1:t},\theta) = p(\mathbf{x}_{1,1:t} \mid \mathbf{x}_{2,1:t}, \mathbf{y}_{1:t},\theta)p(\mathbf{x}_{2,1:t} \mid \mathbf{y}_{1:t},\theta)  \label{rbpfdecomp}.
\end{eqnarray}
If $p(\mathbf{x}_{1,1:t} \mid \mathbf{x}_{2,1:t}, \mathbf{y}_{1:t}, \theta)$ is tractable, and $p(\mathbf{x}_{2,1:t} \mid \mathbf{y}_{1:t}, \theta)$ is not, then samples are only needed for the latter. The particles are comprised of $\{\tilde{w}^i_{1:t}, \mathbf{x}^i_{1,1:t}, \mathbf{x}_{2,1:t}^i \}$, where the $\mathbf{x}^i_{1,1:t}$ are nonrandom summaries of the distribution $p(\mathbf{x}_{1,1:t} \mid \mathbf{x}^i_{2,1:t}, \mathbf{y}_{1:t}, \theta)$, and the $\mathbf{x}_{2,1:t}^i$ are ordinary samples in the sense they represent realizations of the random vector. For example, if the first distribution in (\ref{rbpfdecomp}) is multivariate normal, then each $\mathbf{x}^i_{1,1:t}$ is a sequence of mean vector/covariance matrix pairs coupled with a corresponding $\mathbf{x}^i_{2,1:t}$ sample, and a real-valued weight $\tilde{w}^i_{1:t}$. 

In our model, $\mathbf{y}_{1:T}$ is the collection of the vectors of arithmetic returns, $\mathbf{x}_{1,1:t}$, conditional on $\mathbf{x}_{2,1:t}$, is a finite state Markov chain, and each particle element $\mathbf{x}^i_{1,t}$ is a probability vector. At each time, we choose a proposal density that only targets the second level of the hidden chain. At time $t=1$, we have $q(\mathbf{x}_{2,1} \mid \mathbf{y}_1, \theta)$, and for times $t > 1$, $q(\mathbf{x}_{2,t} \mid \mathbf{x}_{2,t-1}, \mathbf{y}_t, \theta)$. Once we sample from this distribution, we can use a conditional Hidden Markov Model filter, whose conditional likelihoods will be used in the weight update procedure. We then use these weights to resample $n$ indices $a_{t}^{1:n}$ from the multinomial distribution $r(a_{t}^{1:n} \mid \tilde{w}_t^{1:n})$. These indices select the samples we continue to use in following steps. The algorithm is given below in (\ref{alg:RBPF}).

\begin{algorithm}
\caption{RBPF}\label{alg:RBPF}
{\fontsize{7}{2}\selectfont
\begin{algorithmic}[0]
\Procedure{RBPF}{}
  \If{$t$ equals $1$}
    \For{$i=1,\ldots, n$}
      \State Draw $\mathbf{x}_{2,1}^i \sim q_1(\mathbf{x}_{2,1} \mid \mathbf{y}_1, \theta)$  
      \State Compute $p(\mathbf{x}^i_{1,1} \mid \mathbf{x}^i_{2,1}, \mathbf{y}_1, \theta)$ using $p(\mathbf{x}^i_{1,1} \mid \mathbf{x}^i_{2,1}, \theta)$
      \State Compute $p(\mathbf{y}_1 \mid \mathbf{x}_{2,1}^i)$  
      \State Compute $w_1^i = \frac{p(\mathbf{y}_1\mid \mathbf{X}_{2,1}^i, \theta) f(\mathbf{x}_{2,1}^i, \theta)}{q_1(\mathbf{x}_{2,1}^i \mid \mathbf{y}_1, \theta)}$ 
      \State Normalize $\tilde{w}_1^i = w_1^i / \sum_j w_1^j$
    \EndFor
    \State Store $\log\left[\tilde{p}(\mathbf{y}_1, \theta)\right]$ and compute expectation approximations
	\For{$i=1,\ldots, n$}
		\State Draw $a_1^i \sim r( a_1^i \mid \tilde{w}_1^{1:n})$
	\EndFor    
  \Else
  \For{$i=1,\ldots N$}
    \State Draw $\mathbf{x}_{2,t}^i \sim q_t(\mathbf{x}_{2,t} \mid \mathbf{x}_{2,t-1}^{a_{t-1}^i},\mathbf{y}_t, \theta)$
    \State Compute $p(\mathbf{x}^i_{1,t} \mid \mathbf{x}^i_{2,t}, \mathbf{y}_{1:t}, \theta)$
    \State Compute $p(\mathbf{y}_t \mid \mathbf{y}_{1:t-1}, \mathbf{x}_{2,1:t}^i, \theta ) $ 
    \State Compute $w_t^i = w_{t-1}^i * \frac{f(\mathbf{x}^i_{2,t} \mid \mathbf{x}^{a_{t-1}^i}_{2,t-1}, \theta) p(\mathbf{y}_t \mid \mathbf{y}_{1:t-1}, \mathbf{x}^i_{2,1:t}, \theta )}{q_t(\mathbf{x}^i_{2,t}\mid \mathbf{x}^{a_{t-1}^i}_{2,t-1},\mathbf{y}_t, \theta) }$
    \State Normalize $\tilde{w}_t^i = w_t^i / \sum_j w_t^{j}$
  \EndFor
  \State Store $\log\left[\tilde{p}(\mathbf{y}_t \mid \mathbf{y}_{1:t-1})\right]$ and compute expectations
  \For{$i=1,\ldots, n$}
	\State Draw $a_t^i \sim r( a_t^i \mid \tilde{w}_t^{1:n})$
   \EndFor    
  \EndIf
\EndProcedure
\end{algorithmic}
}
\end{algorithm}

If it is desired, approximating expectations and conditional likelihoods should be done before resampling. The conditional likelihood estimates can be computed as follows:
\begin{eqnarray}
\tilde{p}(\mathbf{y}_t \mid \mathbf{y}_{1:t-1}) = \sum_{i=1}^N \tilde{w}_{r+1:t-1}^i \frac{f(\mathbf{x}^i_{2,t} \mid \mathbf{x}^i_{2,t-1}) p(\mathbf{y}_t \mid \mathbf{y}_{1:t-1}, \mathbf{x}^i_{2,1:t} )}{q_t(\mathbf{x}^i_{2,t}\mid \mathbf{x}^i_{2,t-1}, \mathbf{y}_t) },
\end{eqnarray}
and
\begin{eqnarray}
\tilde{p}(\mathbf{y}_1) = \frac{1}{N}\sum_{i=1}^N \frac{p(\mathbf{y}_1\mid \mathbf{x}_{2,1}^i) f(\mathbf{x}_{2,1}^i)}{q_1(\mathbf{x}_{2,1}^i \mid \mathbf{y}_1)},
\end{eqnarray}
while expectations $\mathbb{E}[h(\mathbf{x}_{t}) \mid \mathbf{y}_{1:t}, \theta] = \mathbb{E}[ \mathbb{E}\left\{h(\mathbf{x}_{t}) \mid \mathbf{x}_{2,t}, \mathbf{y}_{1:t}\right\} \mid \mathbf{y}_{1:t}, \theta]$ are approximated as
\begin{eqnarray}
\widehat{\mathbb{E}}[h(\mathbf{x}_t) \mid \mathbf{y}_{1:t}, \theta] = \sum_{i=1}^n \tilde{w}_t^i \mathbb{E}\left(h(\mathbf{x}_{1,t},\mathbf{x}_{2,t}^i) \mid \mathbf{x}_{2,t}^i, \mathbf{y}_{1:t}\right) .
\end{eqnarray}
Note that all of these expressions are free of $\mathbf{x}_{1,t}$. If we were interested in estimating a model that possessed a very large state space for $\mathbf{x}_{1,t}$, we would only need to sample $\mathbf{x}_{2,t}$, which is of much smaller dimension. It is likely that the accuracy of the particle filter's approximations would not suffer; however, this comes at the expense of more computationally intensive computations needed for the closed-form updates.

\subsection{The Particle Marginal Metropolis-Hastings Algorithm}

The Metropolis-Hastings algorithm targets a posterior distribution $p(\theta \mid \mathbf{y}_{1:T})$ if it simulates a Markov chain $\{\theta^i\}$ that possesses as its stationary distribution this posterior of interest \cite{brooks2011handbook}. At each stage in the algorithm, one has a current sample $\theta^i$, and uses that value to sample a proposal $\vartheta \sim q(\vartheta \mid \theta)$. With probability $\min\left(1, \frac{p(\vartheta \mid \mathbf{y}_{1:T}) q(\theta \mid \vartheta) }{p(\theta \mid \mathbf{y}_{1:T}) q(\vartheta \mid \theta) } \right)$ the sample is accepted, which means $\theta^{i+1}$ is set to $\vartheta$. Otherwise, the sample is rejected and $\theta^{i+1}$ is set to $\theta^{i}$. 

To use this algorithm, one must be able to evaluate an unnormalized version of the posterior distribution. Unfortunately, this is often not possible when using factor stochastic volatility models because, as was mentioned before, this marginal likelihood is a high-dimensional integral. This motivates the particle marginal Metropolis-Hastings algorithm (PMMH), which sidesteps this difficulty by introducing a large collection of auxiliary random variables \cite{pmcmc}. This collection of auxiliary random variables, $\mathbf{U}$, is the entire collection of output generated by a particle filter that runs at every iteration of the chain. At every iteration, these samples are used to calculate an estimator of the conditional likelihood $\hat{p}(\mathbf{y}_{1:T} \mid \theta) = h(\mathbf{u},\theta,\mathbf{y}_{1:T})$, which is used as a substitute in the acceptance probability. 

Programatically, everything else about this algorithm is the same as the previous one, which suggests that it targets some approximation to the posterior distribution. Surprisingly, however, this algorithm is asymptotically exact---it targets the distribution of interest exactly. More precisely, the PMMH algorithm targets the following joint posterior:
\begin{eqnarray}
p(\theta, \mathbf{u} \mid \mathbf{y}_{1:T})  = p(\mathbf{u} \mid \theta, \mathbf{y}_{1:T}) p(\theta \mid \mathbf{y}_{1:T})= \frac{\tilde{p}(\mathbf{y}_{1:T} \mid \theta) \psi(\mathbf{u} \mid \theta, \mathbf{y}_{1:T}) }{p(\mathbf{y} \mid \theta)}p(\theta \mid \mathbf{y}_{1:T}),
\end{eqnarray}
where $\psi(\mathbf{u} \mid \theta, \mathbf{y}_{1:T})$ is the distribution of output generated by the particle filter. It is easy to see that as long as the likelihood estimator is unbiased, or in other words, that $\mathbb{E}_{\psi}[\tilde{p}(\mathbf{y}_{1:T} \mid \theta)] = p(\mathbf{y}_{1:T} \mid \theta)$, then the marginal target is the correct posterior. A proof of the RBPF's unbiasedness is given in \cite{mythesis}.

\cite{andrieu2016} (Theorem 10) show that less-dispersed likelihood estimators should be preferred, all else constant. Specifically, one normalized likelihood estimator $\tilde{p}(\mathbf{y}_{1:T} \mid \theta)_1 / p(\mathbf{y}_{1:T} \mid \theta)$ will lead to a smaller asymptotic variance over $\hat{p}(\mathbf{y}_{1:T} \mid \theta)_2 / p(\mathbf{y}_{1:T}\mid \theta)$ when the expectation of any convex function of the former is smaller than the expectation of the latter. Or in other words, when they are ``convex ordered." 

They mention that averaging estimators and using Rao-Blackwellized estimators can be useful strategies if they are available and not computationally prohibitive. The second strategy has also been described by \cite{pmcmcdiscussions} and \cite{rbpfipmmh}. We use both of these strategies in our algorithm. Using more threads to run particle filters in parallel comes at almost no extra computation time, and is only slightly more difficult to program. The Rao-Blackwellization scheme seems obvious, but the name of this filter comes from the local Rao-Blackwellization happening at every time step. It is less obvious, but still true, that the overall likelihood estimator is also Rao-Blackwellized.

\begin{lemma}\label{lemma1}
Let $\hat{p}(\mathbf{y}_{1:T}) = \hat{p}(\mathbf{y}_1)\prod_{t=2}^T\hat{p}(\mathbf{y}_t \mid \mathbf{y}_{1:t-1})$ be the likelihood estimator obtained from a SISR algorithm using the proposal distributions $q(\mathbf{x}_1 \mid \mathbf{y}_1, \theta) = q(\mathbf{x}_{1,1} \mid \mathbf{x}_{2,1}, \mathbf{y}_1, \theta) q(\mathbf{x}_{2,1} \mid \mathbf{y}_1, \theta)$ at time $t=1$, and $q(\mathbf{x}_{t} \mid \mathbf{x}_{t-1}, \mathbf{y}_t, \theta) = q(\mathbf{x}_{1,t} \mid \mathbf{x}_{2,t}, \mathbf{x}_{1,t-1}, \mathbf{y}_t, \theta)q(\mathbf{x}_{2,t} \mid \mathbf{x}_{2,t-1}, \mathbf{y}_t, \theta)$ at times $t > 1$, and where multinomial resampling is conducted at every time point. Also let $\tilde{p}(\mathbf{y}_{1:T}) = \tilde{p}(\mathbf{y}_1)\prod_{t=2}^T\tilde{p}(\mathbf{y}_t \mid \mathbf{y}_{1:t-1})$ be the likelihood estimator obtained from the corresponding Rao-Blackwellized particle filter with multinomial resampling conducted at every time point. Then
\[
\tilde{p}(\mathbf{y}_{1:T}) = E\left[ \hat{p}(\mathbf{y}_{1:T})  \bigg\rvert \mathbf{x}_{2,1:T}^{1:n}, a_{1:T-1}^{1:n} \right],
\]
where $a_{1:T-1}^{1:n}$ are the indices selected by the resampling procedure.
\end{lemma}

Assuming each filter runs through the data in the same amount of time, the RBPF can do no worse. However, if they are not implemented carefully, RBPFs can be much slower than SISR filters. Recall that, in addition to retaining samples and weights at every time point, one must also retain either probability vectors, or mean and covariance information. If $\mathbf{x}_{1,t}$ is high-dimensional and/or the number of particles is large, there can be a substantial increase in cost associated with extra matrix multiplications and storage of these variables in memory. Whenever possible, one must exploit the structure of $\mathbf{x}_{1,t}$.

Our PMMH algorithm is given in (\ref{alg:PMMH}). Note that the outer loop is not shown. 

\begin{algorithm}
\caption{a PMMH algorithm}\label{alg:PMMH}
{\fontsize{8}{2}\selectfont
\begin{algorithmic}[0]
\Procedure{pmmh}{$q_t$}
  \If{$i$ equals $1$}
    \State theta[1] $\gets \theta^1$  
    \State run $n_p$ RBPFs in parallel through data $\mathbf{y}_{1:T}$ each using $\theta^1$ and store each $\tilde{p}^i(\mathbf{y}_{1:T} \mid \theta^1)$ for $i=1,\ldots,n_p$
    \State store $\tilde{p}^*(\mathbf{y}_{1:T} \mid \theta^1) = n_p^{-1} \sum_i \tilde{p}^i(\mathbf{y}_{1:T} \mid \theta^1)$
  \Else
    \State draw $\theta' \sim q_t(\theta' \mid \theta^{i-1})$
    \State $n_p$ RBPFs in parallel through data $\mathbf{y}_{1:T}$ each using $\theta'$ and store each $\tilde{p}^i(\mathbf{y}_{1:T} \mid \theta')$ for $i=1,\ldots,n_p$
    \State store $\tilde{p}^*(\mathbf{y}_{1:T} \mid \theta') = n_p^{-1} \sum_i \tilde{p}^i(\mathbf{y}_{1:T} \mid \theta')$
    \State draw $U \sim \text{Uniform}(0,1]$
    \If{$U < \min\left(1 , \frac{p(\theta')\tilde{p}^*(\mathbf{y}_{1:T} \mid \theta')q_t(\theta^{i-1} \mid \theta')}{p(\theta^{i-1})\tilde{p}^*(\mathbf{y}_{1:T} \mid \theta^{i-1})q_t(\theta' \mid \theta^{i-1})} \right) $}
      \State $\theta^i \gets \theta'$
    \Else
      \State $\theta^i \gets \theta^{i-1}$
    \EndIf
  \EndIf
\EndProcedure
\end{algorithmic}
}
\end{algorithm}

\section{Empirical Studies}

The data set we used for model estimation is comprised of the arithmetic returns for the Select Sector SPDR exchange traded funds. The data from the window spanning 2005-12-30 to 2007-11-23 was used as training data, and data from 2007-11-30 to 2014-05-02 is used as out-of-sample data for the forecast evaluations. Further, we assume $d_f = 1$, and $K=2$. 

First, we describe the prior distributions chosen for the parameters, along with the parameter estimates obtained using our PMMH algorithm on the training data. Then, using these parameter estimates, we test forecasts on out-of-sample data that is a) much longer than the training data, and b) includes the 2008 financial crisis. 

\subsection{Prior Distributions}

We assume all of the parameters are independent in the prior. In other words, our prior is of the following form:
\[
\pi(\theta) = \pi(\mathbf{B})\pi(\bm{\phi})\pi(\bm{\mu})\pi(\bm{\sigma}^2)\pi(\mathbf{R})\pi(\lambda)\pi(p).
\]
We assume each free element of the loadings matrix independently and identically follows a normal distribution. That is $\mathbf{B}_i \sim \text{Normal}(1.0, .125)$ for $i=2, \ldots, 9$. We assume all loadings are close to $1$, because that is a typical estimate one when regresses returns on the observable overall market returns. These types of regression models also go by the name factor models in the literature. However, the factors of these models are always observable returns from constructed portfolios.

We also assume that $\phi_i \sim \text{Uniform}(.4,.9)$. This enforces stationarity of the log volatility process. It also reflects our prior knowledge of the tendency for log volatilities to be positively autocorrelated, so that high-magnitude returns tend to follow more high-magnitude returns in time. This phenomenon is well known and goes by the name of ``volatility clustering." Note that we enforce an upper bound of $.9$. The reason we choose a number less than $1$ is because we use lower-frequency weekly returns, and we suspect our panic factor will help explain some of this "persistence."

It is assumed that the market's risk premium parameter $\lambda \sim \text{Uniform}(1.5\times 10^{-4}, .002708178)$. This noninformative prior corresponds with annual returns between $.783$\% and $15$\%.

Noninformative priors are put on the idiosyncratic variance terms because we will allow for each ETF to be affected by news only pertaining to that particular sector: for $i=1,2$, $R_i \sim \text{InverseGamma}(.001,.001)$. Also each $\mu_i \sim \text{Normal}(0, 1)$ and each $\sigma^2_i \sim \text{InverseGamma}(1,1)$.

\subsection{Estimation Results}

A multivariate normal random walk distribution is used to propose parameters in the PMMH algorithm. The covariance matrix for this algorithm is tuned during the $t_0 = 150$th to the $t_1 = 1000$th iteration of the algorithm using the following formula:

\begin{eqnarray}
\Sigma_i =
\begin{cases}
\Sigma_0 & i \le t_0 \\
\frac{2.4^2}{d} \left[ S_{i-1}^2 + \epsilon I_{d_{\theta}}\right]& t_1 \ge i > t_0 \\
\Sigma_{i-1} & i > t_1,
\end{cases}
\end{eqnarray}
where $S_n$ is the sample covariance matrix of the parameter samples. This formula is identical to that of \cite{haario2001}. Note that the adjustment stops at iteration $1000$, and so standard asymptotic results for MCMC algorithms still apply.

The program is run for approximately $70$ hours to obtain $100,000$ samples using $28$ cores. At each of the $100,000$ iterations of the chain, these $28$ cores each run a Rao-Blackwellized particle filter through the returns data with $50$ particles. The estimates and their associated standard errors for the two models are produced below. More plots can be found in the appendix.

The program, written entirely in \verb|c++|, makes use of two static libraries: \verb|pf| helps to instantiate particle filter objects, and \verb|ssme| has classes that perform the sampling. This particular Gaussian random walk MCMC strategy is handled by the class template \verb|ssme::ada_pmmh_mvn|. The particle filter class for this particular state space model subclasses \verb|pf::rbpf_hmm_bs|. Links to these materials can be found in the supplementary materials. 

The parameter estimates are given in Table \ref{tab:est}. Comparing the two log volatility processes, the one corresponding with the market factor has lower volatility of volatility, is less persistent, but has a higher long-run average value. The values of the elements of $\widehat{\mathbf{B}}$ and $\widehat{\mathbf{R}}$ represent each ETF's riskiness in different ways and appear to be economically reasonable. For example, the energy sector appears to have a high idiosyncratic error variance ($\hat{R}_2 = 6.61963$), and the consumer stamples ETF has a low responsiveness to the market ($\hat{B}_6 = 0.56708$). 

\begin{table}[]
\caption{Parameter Estimates and Uncertainty Measures.  \label{tab:est}}
\centering
\begin{tabular}{rrrrr}
  \hline
 & est & mcse & 95.credible.lower & 95.credible.upper \\ 
  \hline
beta2 & 0.87824 & 0.00623 & 0.59667 & 1.17427 \\ 
  beta3 & 1.02109 & 0.00394 & 0.85015 & 1.20775 \\ 
  beta4 & 0.81926 & 0.00294 & 0.68536 & 0.97437 \\ 
  beta5 & 0.98017 & 0.00401 & 0.78550 & 1.16819 \\ 
  beta6 & 0.56708 & 0.00250 & 0.44181 & 0.69648 \\ 
  beta7 & 0.67692 & 0.00351 & 0.49587 & 0.86416 \\ 
  beta8 & 0.65767 & 0.00274 & 0.52099 & 0.79941 \\ 
  beta9 & 0.90821 & 0.00367 & 0.73534 & 1.07648 \\ 
  phi1 & 0.61626 & 0.00704 & 0.40020 & 0.82090 \\ 
  phi2 & 0.67449 & 0.00768 & 0.42984 & 0.89432 \\ 
  mu1 & 1.00294 & 0.01414 & 0.48041 & 1.56409 \\ 
  mu2 & 0.62478 & 0.02489 & -0.39325 & 1.53487 \\ 
  ss1 & 0.40486 & 0.01419 & 0.10576 & 0.78439 \\ 
  ss2 & 0.53230 & 0.01491 & 0.15228 & 1.11280 \\ 
  R1 & 1.80174 & 0.01590 & 1.22484 & 2.57642 \\ 
  R2 & 6.61963 & 0.03791 & 4.89888 & 8.75874 \\ 
  R3 & 0.83710 & 0.01009 & 0.46230 & 1.27230 \\ 
  R4 & 0.55770 & 0.00530 & 0.31959 & 0.80198 \\ 
  R5 & 1.31760 & 0.01008 & 0.80772 & 1.85621 \\ 
  R6 & 0.65118 & 0.00585 & 0.43209 & 0.90328 \\ 
  R7 & 2.19223 & 0.01610 & 1.43311 & 2.96094 \\ 
  R8 & 0.67863 & 0.00832 & 0.37595 & 1.01083 \\ 
  R9 & 0.38425 & 0.00623 & 0.16684 & 0.63382 \\ 
  lambda1 & 0.00130 & 0.00004 & 0.00035 & 0.00269 \\ 
  p & 0.87132 & 0.00311 & 0.77283 & 0.97320 \\ 
   \hline
\end{tabular}
\end{table}

\subsection{Out-of-sample testing results}


For the out-of-sample forecasting test, we use the expressions for the forecast mean and forecast covariances. The forecast mean is constant and can be written as
\[
\mathbb{E}[\mathbf{y}_{t+1} \mid \mathbf{y}_{1:t}, \theta] = \mathbb{E}[\mathbf{y}_{t+1} \mid x_{2,t}, x_{3,t}, \theta] = \lambda \mathbf{B}.
\]
The forecast covariances have expressions that are more complicated. They can be approximated at each time point using the particle weights owing to the fact that 
\begin{align*}
\mathbb{V}[\mathbf{y}_{t+1} \mid \mathbf{y}_{1:t}, \theta] &= \mathbb{E}[\mathbb{V}\left( \mathbf{y}_{t+1} \mid \mathbf{x}_{2,t}, \theta \right)\mid \mathbf{y}_{1:t}, \theta] + \mathbb{V}[\mathbb{E}\left( \mathbf{y}_{t+1} \mid \mathbf{x}_{2,t}, \theta  \right)\mid \mathbf{y}_{1:t}, \theta] \\
&= \mathbb{E}[\mathbb{V}\left( \mathbf{y}_{t+1} \mid \mathbf{x}_{2,t}, \theta  \right)\mid \mathbf{y}_{1:t}, \theta] \\
&\approx \sum_{i=1}^n \tilde{w}_t^i \mathbb{V}[\mathbf{y}_{t+1} \mid x_{2,t}^i, \theta].
\end{align*}
The conditional variance term in the last line is equal to 
\begin{align*}
\mathbb{V}[\mathbf{y}_{t+1} \mid \mathbf{x}_{2,t}, \theta] &= \exp(\mu_1 + \phi_1(x_{2,t}-\mu_1) +  \sigma_1^2/2) \mathbf{B}\mathbf{B}^{\intercal} \\
&+ \exp(\mu_2 + \phi_2(x_{3,t}-\mu_2) + \sigma_2^2/2) \overline{\mathbf{D}} \overline{\mathbf{B}} \overline{\mathbf{B}}^{\intercal} \overline{\mathbf{D}}
+ \mathbf{R},
\end{align*}
where $\overline{\mathbf{D}} \overline{\mathbf{B}} \overline{\mathbf{B}}^{\intercal} \overline{\mathbf{D}} = \mathbb{E}\left[  \mathbf{D}(x_{1,t+1})  \mathbf{B} \mathbf{B}^{\intercal} \mathbf{D}(x_{1,t+1})\mid \mathbf{x}_{2,t}, \theta \right]$. 

\begin{figure}
\begin{center}
\includegraphics[scale=.65,bb=0 0 504 504]{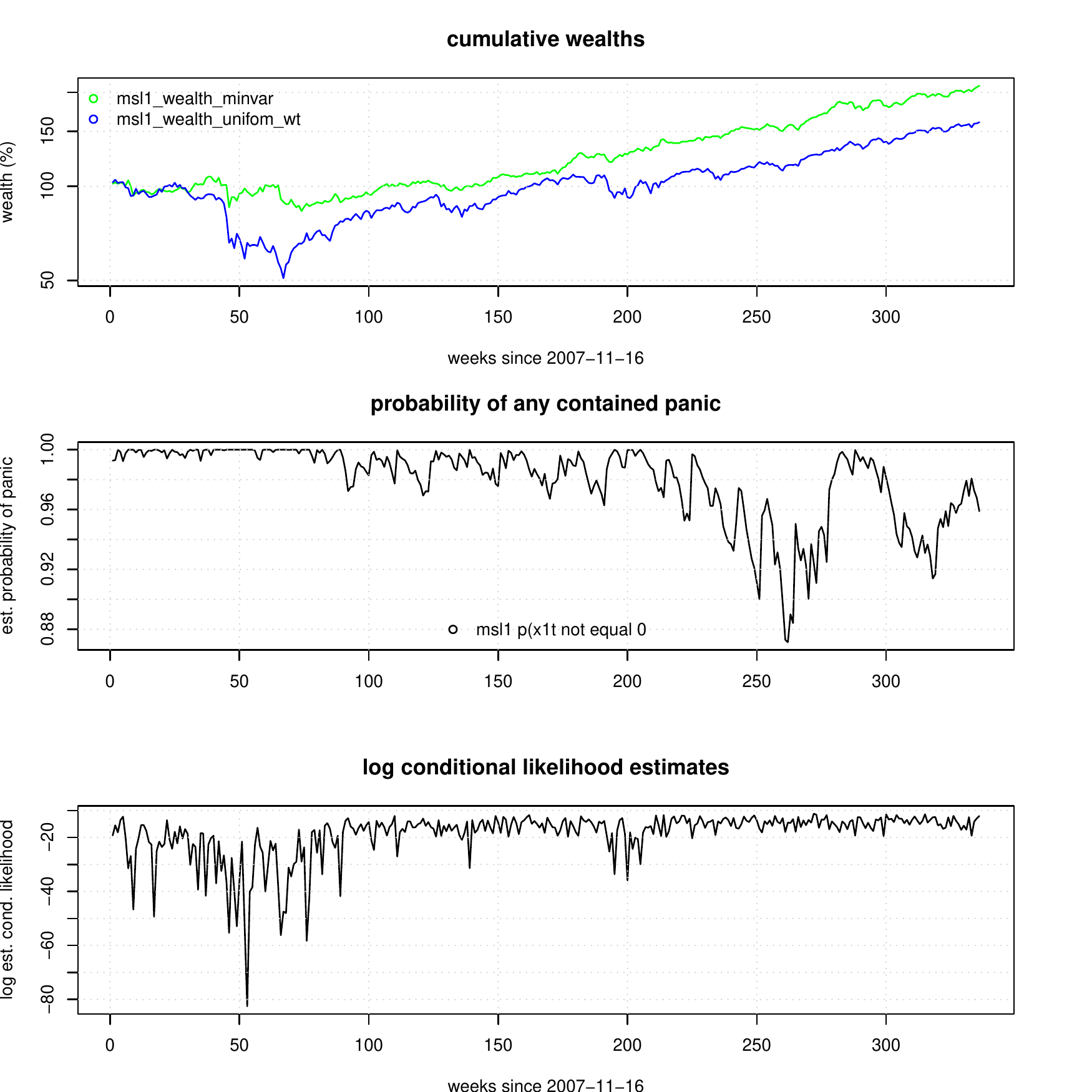}
\end{center}
\caption{Out-of-sample return results for MSL1 (red) and MSL2 (green). The blue curve }\label{fig:first}
\end{figure}

Two tasks are accomplished using these forecast moments. First, results of a hypothetical investing strategy are obtained. Each week this strategy will choose portfolio weights in order to yield an overall return with minimal forecast variance, constraining the weights to sum to unity. Costs such as the impact of trading, commissions, and other fees are ignored. Figure \ref{fig:first} displays the cumulative wealth curves generated by this strategy and the strategy that maintains equal weights. It also displays the filtered probabilities of there being any contained panic $P(x_{1,t} > 0 \mid \mathbf{y}_{1:t})$, and the estimated log conditional probability. The final quantity is used to assess the model's forecasting capability, and can be used to approximate an evaluation of the score function \cite{mythesis}. All results were obtained using one particle filter with $100$ particles instantiated with the estimated posterior means. It is also an option to instantiate many particle filters using as parameters random draws from the MCMC samples.

Second, value at risk (VaR) exceedances are computed. At each week, the (approximate) lower $5$th percentile from the forecast distribution are calculated, and a note is made whether or not the return breaches this lower bound. These quantiles are calculated with the following expression:

\[
\mathbf{w}_t^{\intercal} \hat{\mathbb{E}}[\mathbf{y}_{t+1} \mid \mathbf{y}_{1:t}, \theta] - z_{\alpha}\sqrt{\mathbf{w}_t^{\intercal} \hat{\mathbb{V}}[\mathbf{y}_{t+1} \mid \mathbf{y}_{1:t}, \theta]\mathbf{w}_t  }
\]
where $\mathbf{w}$ is the vector of portfolio weights chosen at the end of time $t$,and $z_{\alpha}$ is the $100(1-\alpha)$th percentile from a standard normal distribution. The financial literature is rich with references detailing how to calculate these quantiles in alternative ways, but it is not explored any further here. It should also be noted that this quantity can be approximated with the output from a particle filter; however, this option is chosen because it is the simplest, and because it recycles the forecast means and covariances that have already been computed.

\begin{figure}
\begin{center}
\includegraphics[scale=.5,bb=0 0 504 504]{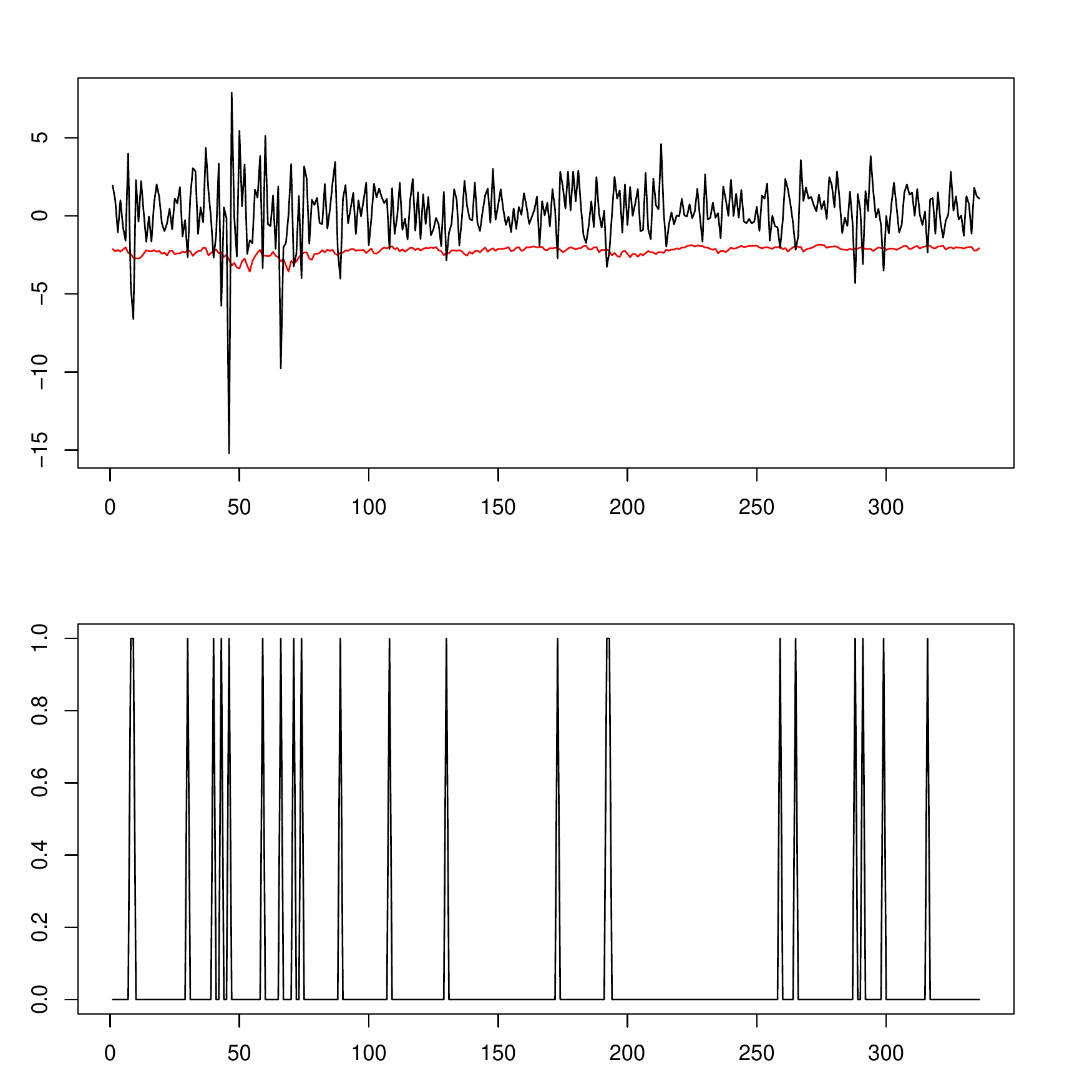}
\end{center}
\caption{Out-of-sample VaR results for MSL1. Top: returns in black, VaRs in red. Bottom: exceedances versus in time.}\label{fig:second}
\end{figure}

Out of $336$ return observations, $22$ (approximately $6.5$ percent) of them exceed the value at risk estimated by the MSL model. Both the unconditional and conditional coverage Value at Risk Exceedances Test fail to reject the null hypothesis \cite{test_var1}, \cite{test_var2}, \cite{rugarch}.

\section{Conclusion}
\label{sec:conc}

A multivariate Markov-switching factor stochastic volatility model has been presented, along with an estimation technique that has the potential to allow users to easily estimate alternative specifications of this model, as well as choose from a wide array of prior distributions. Estimating these models will not require model-specific derivations. The estimation procedure is computationally intensive, but if one distributes the work across more cores, this strategy will become more feasible.

\bigskip
\begin{center}
{\large\bf SUPPLEMENTARY MATERIAL}
\end{center}

\begin{description}

\item[Proof of all Lemmata: ] Proves that the likelihood estimate from a Rao-Blackwellized Particle Filter is a conditional expectation, and proves every lemma about nonidentifiability in the MSL model.

\item[c++ particle filter code: ] The pf static library to be used to instantiate particle filter objects.

\item[c++ particle pmmh code: ] The ssme static library to be used to instantiate estimation objects that automatically handle all required computation.

\item[project code and data: ] Project-specific code and data. Includes the R script that cleans raw data, and c++ code that estimates models, and uses them to forecast the out-of-sample data. 

\item[plots of MCMC samples:] extra mcmc plots such as trace plots, autocorrelation plots, and plots of the running mean.

\end{description}

\bibliographystyle{Chicago}

\bibliography{Bibliography-MM-MC}
\end{document}